\documentstyle [12pt,a4,epsfig] {article}
\newfont{\sans}{cmss10}
\begin{document}
{\large \hskip 11cm DFUB 98/17 }

{\hskip 11cm Bologna, July 1998}
\vskip 0.4cm
{\Large\bf Low energy atmospheric $\nu_\mu$ in MACRO}
\footnote{Contributed paper for the 16th ECRS - Alcala de Henares (Madrid) -
20/25 July 1998}
\vskip 0.4 cm
{\large M. SPURIO (For the MACRO collaboration)}
\vskip 0.1 cm
{\it Universit\`a di Bologna and INFN - Viale Berti Pichat 6/2 -40127 Bologna.}
\vskip 0.1 cm
{\it Email: spurio@bo.infn.it}

\vskip 0.8 cm

{\bf Abstract}

{\small The flux of low energy neutrinos 
($\overline E_\nu \sim 4\ GeV$) has been studied with the MACRO 
detector at Gran Sasso via
the detection of $\nu_\mu$ interactions inside the apparatus, and of
upward-going stopping muons.
Data collected in $\sim 3\ y$ with the full apparatus were analyzed.
The results are compatible with 
a  deficit of the flux of atmospheric $\nu_\mu$
from below, and no reduction from above, with respect to Monte Carlo 
predictions. The deficit and the angular distributions are interpreted in 
terms of neutrino oscillations, and compared with the MACRO
results on the upward throughgoing muons ($\overline E_\nu \sim 100\ GeV$) .}

\vskip 0.8 cm
{\bf Introduction}\vskip 0.2cm

Recent results from Super-Kamiokande (Fukuda 98) confirmed the
anomaly in the ratio of contained $\nu_\mu$ to $\nu_e$ interactions, 
suggesting best fit parameters of
$\sin^2 2\theta \simeq 1.0$ and $\Delta m^2$ in the range of a few times
10$^{-3}$ eV$^2$ for  $\nu_\mu$ disappearance. 
Other positive results come from Kamiokande  (Fukuda 94),
IMB (Casper 91) and Soudan 2 (Allison 97) while
earlier results from Frejus (Daum 95) and NUSEX
(Aglietta 1989) are consistent with 
the expected number of contained events with smaller statistics.
Also the results from MACRO (Ahlen 95) on up throughgoing muons
presented an anomaly.

\begin{figure}[tbh]
\epsfig{figure=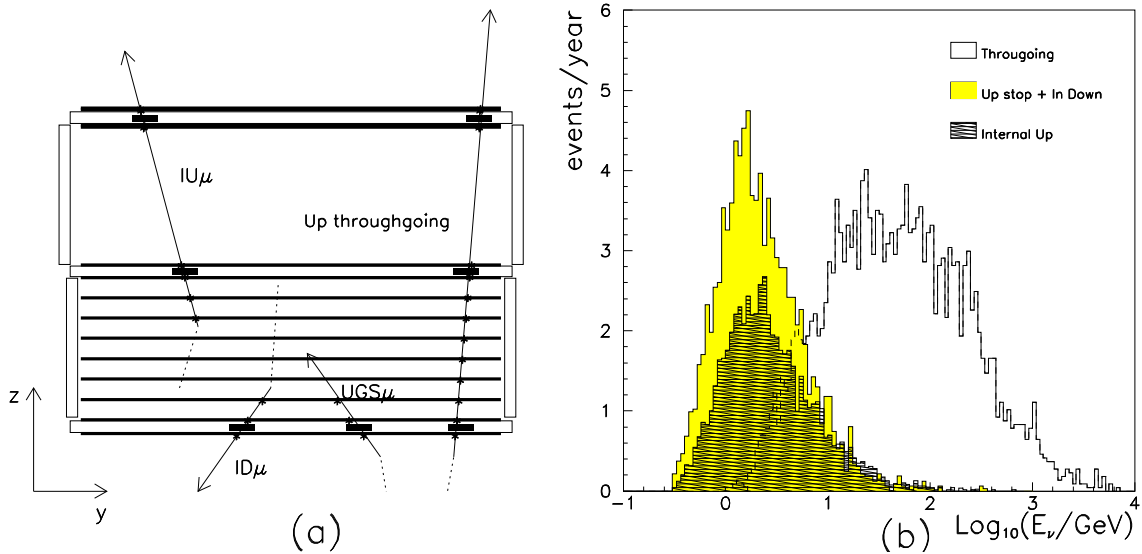,height=8cm}
{\caption {\label{fig:topo}\small (a) Sketch of 
different event topologies induced
by neutrino interactions in or around MACRO.
$IU\mu$= Internal Upgoing $\mu$; 
$ID\mu$= Internal Downgoing $\mu$; 
$UGS\mu$= Upgoing Stopping $\mu$; 
Up throughgoing = upward throughgoing $\mu$.
In the figure, the stars
represent the streamer tube hits, and the black boxes the scintillator hits.
The {\it time-of-flight} of the particle can be measured for the
$IU\mu$ and up throughgoing events. (b) Parent neutrino energy 
distributions for the three $\nu_\mu$ samples illustrated in (a).}}
\end{figure}

The MACRO detector (Ahlen 93)
is a large rectangular box (76.6~m~$\times$~12~m~$\times$~9.3~m)
whose active detection elements are planes of streamer tubes for tracking
and liquid scintillation counters for fast timing. 
 The lower half of the detector is filled
with trays of crushed rock absorber alternating with streamer tube
planes, while the upper part is open.
The low energy $\nu_\mu$ flux can be studied by the
detection of $\nu_\mu$ interactions inside the apparatus, and
by the detection of upward going muons 
produced in the rock below the detector and stopping inside the detector
(Fig.~\ref{fig:topo}a).
Because of the MACRO geometry, muons induced by neutrinos with the interaction
vertex inside the apparatus can be tagged
with {\it time-of-flight} ($T.o.F.$) measurement only for upgoing muons
({\it $IU\mu$=Internal Upgoing $\mu$}).
The downgoing muons with vertex in MACRO
({\it $ID\mu$=Internal Downgoing $\mu$})
and upward going muons
stopping inside the lower part of the 
detector ({\it $UGS\mu$ = Upward Going  Stopping $\mu$})
can be identified via topological constraints.
Fig.~\ref{fig:topo}b shows
the parent neutrino energy distribution for the three event
topologies that can be  detected by MACRO.

\vskip 0.6 cm
{\bf Internal Upgoing Events (IU)}\vskip 0.2cm

The data sample used for the Internal Upgoing (IU) events corresponds to a 
live-time of
3.16 years from April, 1994 up to November, 1997. 
During this period, $\sim 22\ 10^6$ atmospheric single muons were collected.
The identification of IU$\mu$ events was based both on topological
criteria and $T.o.F.$ measurements.
The basic requirement is the presence of at least two
scintillator clusters in the upper part of the apparatus (see Fig.
\ref{fig:topo}a) matching
a streamer tube track reconstructed in space. A similar  
request was made in the analysis for the up throughgoing events (Ambrosio 98).

\begin{figure}
\epsfig{file=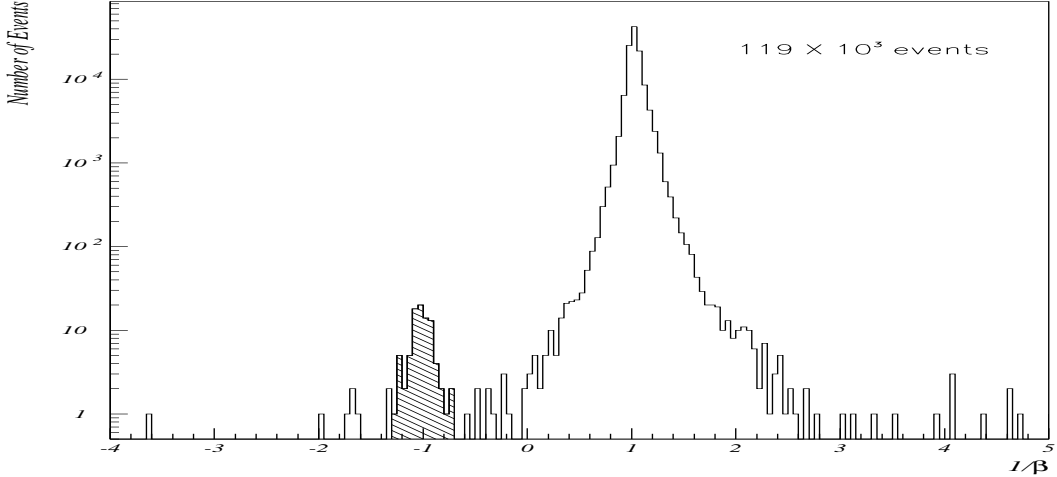,height=6.5cm,width=13cm}
\caption{\label{fig:sbeta}\small
The $1/\beta$ distribution of the 85 IU events (dashed area).
The $\sim 1.2\ 10^5$ events centered at $1/\beta=+1$ 
are downgoing atmospheric stopping muons.}
\end{figure}

For IU$\mu$ candidates, the track starting point must be inside the 
apparatus. To reject fake semi-contained events entering from a 
detector crack, the extrapolation of the muon track in the lower part 
of the detector must cross and not fire 
a minimum number of streamer tube planes and scintillator
counters depending on track configuration. 
The above conditions, tuned with Monte Carlo simulated events,
account for detector inefficiencies and reduce
the background contribution from upward throughgoing muons, which appear
like semi-contained event to $\sim 1 \%$.
The measured $1/\beta$ distribution is shown in Fig.~\ref{fig:sbeta}.
The measured muon velocity $\beta c$ is calculated with the
convention that downgoing muons have
1/$\beta$ values centered at +1 while upgoing muons have 1/$\beta$ at -1.
It was evaluated that three events are due to an uncorrelated background.
After background subtraction, 85 events are classified as IU events.

\vskip 0.6 cm
{\bf UpGoing Stopping (UGS) and Internal Downgoing (ID) Events}
\vskip 0.2cm

The UGS+ID events were identified via topological constraints, and not
with the $T.o.F$. For this analysis, the effective live-time is $2.81\ y$.
The main request for the event selection is
the presence of one reconstructed track crossing
the bottom layer of the scintillation counters (see Fig. \ref{fig:topo}a).
All the hits along the track must be at least one meter inside from
each wall of a MACRO supermodule. 
The selection conditions for the event vertex (or $\mu$ stop point)
to be inside in the detector are similar to
those used for the IU search; they reduce to a negligible level
the probability that an atmospheric muon produces a background event.
The main difference with respect to the IU analysis (apart the $T.o.F.$) 
is that on average fewer streamer tube hits are present.
To reject ambiguous and/or wrongly tracked events
which passed the event selection, a systematic scan with the
MACRO Event Display was performed. 
All real and simulated events which passed the event selection criteria
were randomly merged. The finally accepted events had to
pass the double scan procedure
(the differences are included in the systematic uncertainty). 
In real events, three different subsamples were considered,
according to the number of streamer tube hits, {\it =NWP} 
along the reconstructed track.

\begin{table}[tbh]
\begin {center}
\begin{tabular}
{|c||c|c|c||c|c||c|}\hline
 &  $\overline E_\nu$ & $R_{min}$ & $\nu_\mu$ & Events & Bck. & MC \\
 & $(GeV)$       & $(g\ cm^{-2})$ & $\%$      &  & ($\pm20$\%)& (no osc.) \\ \hline
A (All events)   & 3.1 & 30 &  81 & 241    & 22  & 256       \\
B ($NWP\ge 3$)   & 3.9 &100 &  90 & 125    &  5  & 159       \\
C ($NWP\ge 4$)   & 4.9 &160 &  94 &  66    &  1  &  95      \\ \hline
\end{tabular}
\end {center}
\caption {\small Summary results (see text) for the 
three subsamples of ID+UGS events. }
\label{tab:nenc}
\end{table}

As shown in Table 1, each subsample is distinguished by a different
average parent neutrino energy (col. 2);
minimum range $(g\ cm^{-2})$ of detector material to be
crossed by the neutrino-induced particle (col. 3); 
percentage of $\nu_\mu$ C.C. interactions (col. 4). In 
column 5 are given the  numbers of real events.

The main background source is due to upward going charged particles
(mainly pions) induced by interactions of atmospheric muons in the rock
around the detector (Ambrosio 98).
The background affects mainly sample $A$, and becomes 
smallest in sample $C$, as shown in col. 6 of Table 1.
Sample $B$ is used in the next section.

\vskip 1.0 cm
{\bf Comparisons between Data and Monte Carlo}
\vskip 0.2cm

The expected rates of IU and ID+UGS events were evaluated with a full 
Monte Carlo simulation.
The $\nu_e$ and $\nu_\mu$ were allowed to interact in a large volume of 
rock containing the experimental Hall B
and the detector.  The rock and detector mass in the generation volume
is $175\ kton$.
The atmospheric $\nu_\mu$ flux given by the Bartol group
(Agrawal 96) and  the $\nu$ cross sections of (Lipari 94) were used. 
The detector response was simulated
using GEANT, and simulated events are processed in the
same analysis chain as the real data. In the simulation, the parameters of
the streamer tube and scintillator systems have been chosen in order to
reproduce the real average efficiencies.
The total theoretical uncertainty 
for the muon production from $\nu_\mu$ at these energies
is $\sim 25\%$. The
systematic error  is of the order of $10\%$, 
arising mainly from the simulation of detector response, 
data taking  conditions, analysis algorithm efficiency, 
mass and acceptance of the detector.
With our full MC, 
the prediction for IU events is $144 \pm 36_{theor} \pm 12_{syst}$, while
the  observed events are $85 \pm 9_{stat}$. The  ratio 
$R=(DATA / MC)_{IU} =0.59 \pm 0.06_{stat} \pm 0.15_{theor} \pm 0.06_{syst}$.

\begin{figure}[tbh]
\epsfig{file=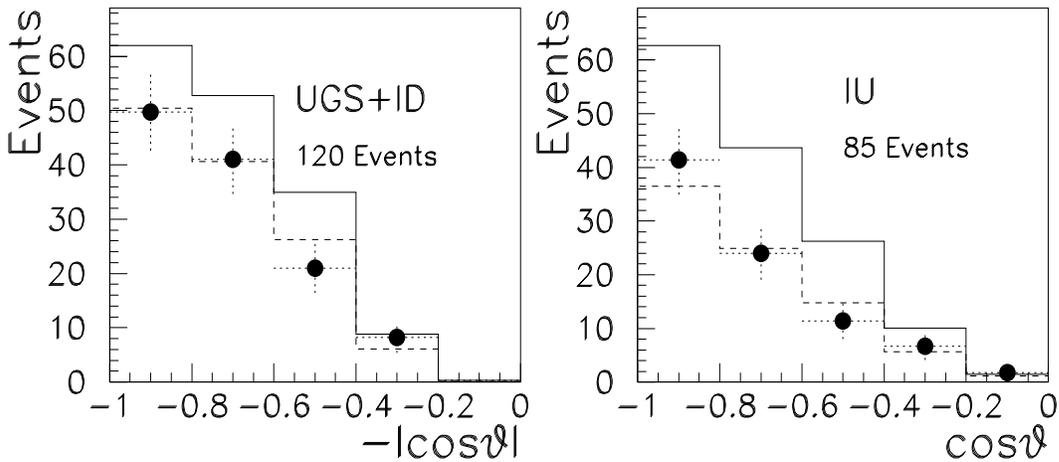,height=6.9cm,width=13cm}
\caption{\label{fig:bole}\small
Distribution of the cosine of the
zenith angle ($\theta$) for ID+UGS (samples B) and IU events.
The background-corrected data points (black points with error bars)
are compared with the Monte Carlo
expectation assuming no oscillation (full line) and two-flavour
oscillation (dashed line) using maximum mixing and $\Delta m^2=2.5\ 10^{-3}\
eV^2$.}
\end{figure}

The MC prediction for ID+UGS events are shown in the last column of Table 1;
in particular for sample B we expect $159\pm 40_{theor} \pm 13_{syst}$ events.
The ratio $R=(DATA / MC)_{ID+UGS}=
0.75 \pm 0.07_{stat} \pm 0.19_{theor} \pm 0.08_{syst}$.
An almost equal number of ID and UGS neutrino induced
events are expected in our data sample.
The angular distribution of the IU and ID+UGS data samples, 
with the Monte Carlo predictions, are presented in Fig.\ref{fig:bole}.

The low energy $\nu_\mu$ samples show a deficit of the measured number of
events  in a uniform way over the whole angular distribution
with respect to the predictions based on the absence
of neutrino oscillations. 
The measured deficit of low-energy events is in agreement with 
the MACRO results on the throughgoing events (Ambrosio 98),
{\it i.e.} with a model of $\nu_\mu$ disappearance
with $\sin^2 2\theta \simeq 1.0$ and
$\Delta m^2 \sim 2.5\ 10^{-3}\ eV^2$. 
In fact, the IU and UGS events have $L\sim 13000\ km$, and in the
energy range of few $GeV$ the flux is reduced by a factor of two for maximum
mixing and $\Delta m^2 \sim 10^{-2} \div  10^{-3}\ eV^2$. No flux reduction
is instead expected for ID events ($L\sim 20 \ km$).
A global analysis  of the different data sets is in progress.

\vskip 0.6 cm
{\bf References}
\vskip 0.2cm

M. Aglietta {\it et al.}, Europhys. Lett. {<\bf 8} (1989) 611.

V. Agrawal, T.K. Gaisser, P. Lipari and T. Stanev, Phys.Rev. {\bf D53} 
(1996) 1314.

S. Ahlen {\it et al.}, (The MACRO coll.) Nucl. Inst. and Meth. {\bf A324} 
(1993) 337.

S. Ahlen {\it et al.}, (The MACRO coll.) Phys. Lett.  {\bf B357} (1995) 481 .

W. Allison  {\it et al.}, Phys. Lett.  {\bf B391} (1997) 491 .

M. Ambrosio {\it et al.}, (The MACRO coll.), INFN-AE 97/55.

M. Ambrosio {\it et al.} (The MACRO coll.), hep-ex/9807005 (1998).

D. Casper {\it et al.}, Phys. Rev. Lett. {\bf 66}  (1991) 2561.

K. Daum {\it et al.}, Z. Phys. C. {\bf 66} 417.

Y. Fukuda {\it et al.}, (Kamiokande Collaboration),Phys. Lett. {\bf B335} (1994) 237.

Y. Fukuda {\it et al.} (Super Kamiokande Collaboration), hep-ex/9807003 (1998).

P. Lipari, M. Lusignoli and F. Sartogo, Phys. Rev. Lett.{\bf 74} (1995) 4384.

\end{document}